\documentclass[sigconf]{acmart}

\AtBeginDocument{%
  \providecommand\BibTeX{{\normalfont B\kern-0.5em{\scshape i\kern-0.25em b}\kern-0.8em\TeX}}}

\copyrightyear{2027}
\acmYear{2027}
\setcopyright{none}
\acmConference[FPGA '27]{The 2027 ACM/SIGDA International Symposium on
  Field-Programmable Gate Arrays}{March 14--16, 2027}{California, USA}
\acmBooktitle{The 2027 ACM/SIGDA International Symposium on Field-Programmable
  Gate Arrays (FPGA '27), March 14--16, 2027, California, USA}
\acmDOI{}
\acmISBN{}

\begin{document}

\title{A State Augmented Approach to Paralle Modeling in High-Level Synthesis}

\author{Nicholas Kovacs}
\affiliation{%
  \institution{University of South Florida}
  \department{Bellini College of Cybersecurity, AI, and Computing}
  \city{Tampa}
  \state{Florida}
  \country{USA}}
\email{imre@usf.edu}

\author{Hao Zheng}
\affiliation{%
  \institution{University of South Florida}
  \department{Bellini College of Cybersecurity, AI, and Computing}
  \city{Tampa}
  \state{Florida}
  \country{USA}}
\email{haozheng@usf.edu}

\begin{abstract}
High-level synthesis (HLS) can schedule only the concurrency made legal by its input description. For feedback-coupled designs, a sequential behavioral model can obscure the architectural state boundaries and hazard resolution needed for efficient overlap. We develop an explicit architectural modeling methodology based on synchronous sequential processes that start in lockstep and synchronize completion before repeating. Persistent state, communication, and feedback control are represented in C++, while HLS schedules and binds operations within that structure. A five-stage RV32I processor illustrates three consequential modeling decisions: which version of a value a consumer reads, when it consumes that value, and how architectural memory actions map to port operations. On a Kintex UltraScale+ FPGA, the CPI-optimized configuration achieves 1.221 cycles per instruction at 220 MHz, compared with 4.000 for an algorithmic HLS baseline and 1.575 at 242 MHz for an Ibex reference. An area-optimized configuration reduces LUTs by 65\% and registers by 47\%, with 9.7\% more benchmark cycles and a 250 MHz achieved clock. These results demonstrate a separation between designer-controlled architecture and tool-controlled scheduling; broader applicability beyond the processor case study remains to be evaluated.
\end{abstract}



\maketitle

\section{Introduction}
High-level synthesis translates sequential source code into parallel hardware through dependence analysis, scheduling, and resource binding. For regular array computations, loop pipelining, unrolling, and task decomposition expose substantial concurrency. Feedback-coupled designs present a different problem: the next computation depends on evolving shared state, and efficient execution requires explicit rules for forwarding, stalls, and redirection. A scheduling directive requests overlap; it does not specify these architectural rules.

A processor makes this distinction concrete. An instruction-set-simulator-style loop fetches, decodes, executes, accesses memory, and commits one instruction. Register indices and branch outcomes are runtime values. Overlapping instructions therefore requires a microarchitecture that handles both value dependencies and control changes. HLS may overlap portions of the behavioral loop, but its legal schedule is constrained by the description's recurrences and timing. In our baseline, the achieved initiation interval (II) is four; forcing a more aggressive schedule fails the clock target.

The proposed methodology represents the design as synchronous sequential processes over explicitly declared architectural boundaries. Processes start in lockstep, compute concurrently subject to communication dependencies, and synchronize completion before committing selected inter-process state and repeating. The designer specifies the communication and consumption timing that make overlap correct; HLS retains detailed scheduling, internal pipelining, resource sharing, and storage binding subject to those semantics. This is architectural modeling in C++, rather than an expectation that directives will reconstruct an unspecified architecture.

Explicit pipeline-state modeling has precedent, particularly in Comet~\cite{comet_iccad,comet}. Our contribution is to develop and experimentally examine the modeling choices that determine cost within this formulation. We isolate three choices: registered versus in-flight communication, the stage at which a value is consumed, and the memory operations required by an architectural action. A complete RV32I case study relates these choices to II, achievable frequency, and area, using an algorithmic HLS baseline, an Ibex RTL reference, and CPI- and area-optimized configurations. The general principle applies to describing feedback-coupled concurrency; the experiments establish its effectiveness for this processor family and tool flow, rather than a universal advantage over RTL or other HLS approaches.

\section{Background and Positioning}
\subsection{Scheduling, State, and Directives}
Loop pipelining overlaps iterations subject to data recurrences, timing, and resource limits. Its II is the interval between iteration starts, not instruction CPI. Dataflow decomposition instead creates communicating tasks, typically with FIFO or channel interfaces. Both mechanisms are useful, but a source-level recurrence must still respect when its state becomes available. True dependencies cannot be removed safely by suppressing dependence checks; false aliasing dependencies may be refined only when independence is established.

The distinction matters for a processor: forwarding sends results backward relative to instruction flow, redirects change the front end, and interlocks selectively hold state. The attempted stream-based formulation reported here makes this concrete. Fetch, decode, execute, memory, and writeback were expressed as free-running \texttt{hls::task} processes connected by streams, with the register file and the completion flag remaining shared state. Vitis reports merging decode, writeback, memory, and execute ``due to feedback on \texttt{regfile}'', then merging all five ``due to feedback on \texttt{is\_finished}'', and extracts a single process function; synthesis then fails because that merged process has non-FIFO I/O. This observation concerns that formulation and tool flow; it does not establish that channel-based designs cannot express feedback.

\subsection{Processor Descriptions and HLS Cores}
Rocket Chip~\cite{rocketchip} and SiFive's E31~\cite{sifive-e31} illustrate generator-based architectural descriptions, while Ibex~\cite{ibex}, used in OpenTitan~\cite{opentitan}, provides a configurable SystemVerilog implementation. PicoRV32~\cite{picorv32} emphasizes compact multi-cycle execution, and VexRiscv~\cite{vexriscv} exposes architectural configuration through SpinalHDL. These designs already state hardware structure rather than asking a software scheduler to infer it. The comparative survey of application-class RISC-V implementations~\cite{riscv_survey} further illustrates why results depend on configuration and implementation context. Here, Ibex is an implementation reference, not an iso-microarchitecture baseline.

Comet~\cite{comet_iccad,comet} is the closest methodological precedent. It compiles one C++ description both into a cycle-accurate simulator and into hardware. Its analysis identifies register-file read-after-write and next-PC dependencies in instruction-level behavioral models. Its pipelined formulation declares persistent inter-stage structs, computes temporary stage outputs, and commits them after evaluating stalls. This gives C++ the current-state/next-state discipline familiar from RTL nonblocking assignments. Forwarding selects available results from later-stage registers, and explicit state machines accommodate caches and multi-cycle operations. Comet reports II=1, but Dhrystone CPI=1.9: static scheduling throughput and runtime stall cost are different quantities.

We share Comet's explicit-state principle and do not claim its invention. Comet establishes that a C++ description can carry persistent inter-stage state under a current-state/next-state discipline; the present study asks what determines performance and area once that structure already exists, and treats the remaining modeling decisions as the object of study rather than as implementation detail. Three such decisions prove consequential, and each changes a measured outcome while leaving the stage structure untouched: which version of a value a consumer reads sets the recurrence delay, when it consumes that value determines which storage bindings are legal, and how an architectural action maps to port operations determines how many accesses must be scheduled. The register file is the clearest example. With branches resolved in decode, the operand read must complete within the same modeled transition, so only a fully partitioned flip-flop register file is legal, and the design pays 601 LUTs and 746 registers for it; moving branch resolution to execute lets the same read cross a stage boundary, after which the identical storage-binding request is accepted as distributed RAM at II=1. The source change is one stage assignment rather than a directive, and no directive alone can produce it. Our fixed-latency evaluation and machine-mode support also differ from Comet's reported configuration, so its Catapult/Artix-7 results provide context, not a controlled cross-tool or CPI comparison.

The method also invites comparison with RTL design, since both describe behavior in terms of cycles and explicit state. The difference is what the description can be used for. The same C++ that is synthesized is compiled by a standard host compiler, so one source serves both as the executable model used to run programs during development and as the input to synthesis; a change to the hazard rules can be exercised in software before any RTL exists, which shortens the loop between an architectural change and functional evidence about it. The correspondence is not automatic: a modeled iteration equals one hardware cycle only when the schedule reaches II=1 and memory latency is fixed, which is why the hardware cycle counts reported here come from an external counter rather than from the model's own iteration count. The resulting quality is competitive with a manually developed core of similar configuration: against Ibex, the area-optimized configuration uses 1,980 LUTs against 1,940 and 1,595 registers against 702, and completes the kernel suite in 0.82 times Ibex's time.

HL5~\cite{hl5} demonstrates a Stratus-generated RV32IM processor and multiple RTL configurations obtained through HLS settings; the scaling study~\cite{hl5_scaling} extends this direction. They establish that settings remain useful once a synthesizable processor description exists. Toker's RV32I SoC~\cite{toker_rv32i} connects HLS simulation, RTL simulation, and FPGA validation. These efforts motivate HLS processor construction and validation. Our emphasis is the causal link between explicit source architecture, legal scheduling, and measured tradeoffs, rather than processor synthesis alone.

\subsection{Directive Search and Structural Transformation}
Automatic pragma insertion~\cite{pragma_nlp} optimizes directive choices using analytical models, while AutoScaleDSE~\cite{autoscaledse} explores transformations and directives through ScaleHLS/MLIR. Such search complements our method: it explores a configuration space whose available concurrency depends on the source formulation.

Structural transformation is also an established research direction. Allo~\cite{allo} separates algorithmic description from composable scheduling transformations. HIDA~\cite{hida} represents functional and structural dataflow hierarchically. Stream-HLS~\cite{streamhls} optimizes loop organization and buffer-to-FIFO communication for affine computations. Sisyphus~\cite{sisyphus} and Prometheus~\cite{prometheus} jointly optimize code transformations, task organization, memory partitioning, and pragmas. These works demonstrate that HLS optimization extends beyond directive tuning; architectural structure is not exclusively a manual concern. Their reviewed evaluations concern regular accelerator computations rather than the runtime-indexed, feedback-coupled processor state examined here. We therefore position explicit architectural modeling as a complementary formulation for this design class, without claiming that automated restructuring is impossible in general.

\section{Explicit Architectural Modeling}
\label{sec:state}
\subsection{Synchronous Sequential Process Model}
The methodology models a design as a set of synchronous sequential processes. Each process describes a portion of the design through sequential computations, while the processes execute concurrently within a common iteration. All processes start in lockstep, and all must complete their work before any process can repeat in the next iteration. Early-finishing processes wait logically at the common completion boundary.

Let $S_t$ denote persistent design state at the start of iteration $t$, and let $P_i$ denote process $i$. Each process computes candidate outputs and updates, with explicitly specified communication, and the synchronized update advances the design:
\begin{equation}
\widehat S_{i,t+1}=P_i(S_t,I_t),\qquad
S_{t+1}=C(S_t,\{\widehat S_{i,t+1}\}_i,H_t).
\end{equation}
Here $C$ selects updates using hold, validity, and redirect conditions $H_t$. Registered communication uses state established at the preceding boundary. Any same-iteration communication or shared-state access must have explicitly defined dependencies and ordering; concurrency does not imply that the processes are independent.

In C++, persistent state is declared outside a top-level loop, and process computations are represented within its body. Candidate outputs are kept separate from current inter-process state until the synchronized update. HLS overlaps process computations where dependencies permit, despite sequential source-level function calls. The enclosing iteration represents completion synchronization, realized by the schedule rather than a separate runtime barrier controller.

The designer specifies process boundaries, sequential behavior within each process, communication, and the synchronized update rules. HLS schedules the operations and binds resources subject to those semantics. Architectural state registers and HLS-inserted scheduling registers have different roles: the former define communication between modeled iterations, while the latter realize the computations under timing and resource constraints. A modeled iteration may occupy multiple hardware cycles; II=1 is a measured scheduling outcome, not an assumption of the process model.

\subsection{RISC-V Example: Five Synchronized Processes}
The RV32I example assigns fetch (IF), decode (ID), execute (EX), memory (MEM), and writeback (WB) to five synchronous sequential processes. Each performs the work of one pipeline stage. During an iteration, the processes operate on different in-flight instructions, start together, and synchronize before repeating. This concurrent execution across processes is distinct from pipelining successive iterations of an instruction-level behavioral loop.

Persistent state comprises the PC, 32-entry 32-bit register file, machine-mode CSRs, and four valid-tagged inter-process registers. IF/ID holds the instruction and its PC; ID/EX holds the decoded operation, register indices, immediate, operand values, and PC; EX/MEM carries the result or memory address, destination, memory controls, and trap information; MEM/WB carries the result and register-write controls. Valid bits identify instructions and bubbles. Reset empties the pipeline, and $x0$ remains unwritable.

Within each iteration, IF reads instruction memory and forms an IF/ID candidate. ID decodes current IF/ID, reads operands, and forms an ID/EX candidate. EX selects available forwarded operands and computes an EX/MEM candidate. MEM performs the access identified by current EX/MEM and forms a MEM/WB candidate. WB commits the valid oldest result. HLS schedules their concurrency while preserving internal dependencies. The source explicitly defines writeback-to-decode visibility for shared register-file accesses; the current-state/next-state discipline applies to the inter-process registers rather than requiring every side effect to be a pure function.

After all process computations and feedback decisions are complete, the synchronized update selects the next PC and commits, holds, or invalidates the inter-process registers. A hazard-induced hold does not let one process advance to another iteration independently: every process still participates in the common iteration, but its selected architectural state may remain unchanged. Fig.~\ref{alg:model} illustrates this execution model; detailed instruction, CSR, and trap logic is omitted.

\begin{figure}[t]
\fbox{\begin{minipage}{0.94\columnwidth}\small
\textbf{Persistent state:} PC, RF, CSRs, IF/ID, ID/EX, EX/MEM, and MEM/WB. Initialize stage registers as bubbles.
\begin{enumerate}\setlength{\itemsep}{0pt}\setlength{\parskip}{0pt}\setlength{\parsep}{0pt}
\item \textbf{Start iteration:} activate all five processes with current inter-process state.
\item \textbf{Concurrent process work:} IF fetches; ID decodes and reads operands; EX forwards and executes; MEM accesses data memory; WB commits a valid result. Respect explicit shared-state and communication dependencies.
\item \textbf{Compute feedback:} evaluate load-use and decode operand interlocks, branch/jump decisions, and trap/MRET redirects.
\item \textbf{Synchronize completion:} finish all process computations before advancing the iteration.
\item \textbf{Select next state:} an older redirect squashes younger work; a front-end interlock holds PC and IF/ID and inserts an ID/EX bubble; otherwise select normal advancement or a decode redirect.
\item \textbf{Commit and repeat:} update selected inter-process state and begin the next common iteration.
\end{enumerate}
\end{minipage}}
\caption{RV32I modeled as synchronized sequential processes. The list specifies phases and concurrent work, not a serial hardware schedule.}
\Description{A six-step list describing one modeled iteration: start all five processes, perform concurrent stage work, compute feedback, synchronize completion, select next state, and commit.}
\label{alg:model}
\end{figure}

\subsection{Feedback, Hazards, and Communication Timing}
For each source operand, forwarding compares its register index with valid later-stage destinations and selects an available matching result instead of the stale decoded value. A load-use dependence cannot use a result before the memory access completes; the interlock holds PC and IF/ID while injecting a bubble into ID/EX, allowing older instructions to advance. Decode-stage branch resolution also requires available operands: the CPI-optimized configuration stalls when a needed operand remains in flight. Taken branches and jumps redirect fetch and invalidate younger wrong-path work. Traps and MRET introduce older redirects from execute; incorrect-path execution and architectural side effects must be suppressed.

Forwarding must prefer the youngest older producer when several in-flight instructions target the same register: selection takes an available EX/MEM result on a valid write match, otherwise a matching MEM/WB result, otherwise the decoded operand, excluding matches to $x0$ and treating a load destination as not yet available. The load-use interlock asserts when ID/EX is a valid load whose nonzero destination matches a source the decoded instruction actually uses. Holding the front end retains the dependent instruction while bubbling ID/EX prevents it from executing with unavailable data, and older stages continue until forwarding or register-file visibility satisfies the dependence.

Control feedback requires a corresponding age discipline. A redirect from an older instruction supersedes younger candidate work, while a decode-resolved branch must wait for its own operands. Validity therefore controls not only register propagation but also whether a candidate may write memory, update a CSR, or retire. PC and instruction metadata travel with in-flight work so that a trap records the faulting instruction rather than the current fetch address. These explicit rules are the architectural content that distinguishes the model from simply splitting sequential instruction semantics into five function calls.

The source distinguishes communication through registered stage state from consuming a newly computed in-flight value. A same-transition bypass can shorten the number of modeled transitions to a consumer, but extends a combinational path and potentially a loop-carried recurrence. A registered path breaks that path at the expense of delayed availability. The designer must specify which result is valid at the consumer and how an unavailable result causes a stall; HLS cannot infer a new hazard protocol from a timing directive. Conversely, making these rules explicit does not remove real dependencies or exempt the model from dependence analysis.

The core includes Zicsr, machine-mode handling of environment calls, breakpoints, illegal instructions, and misaligned loads/stores, and the host-target interface used to run tests. The implementation provides both an AXI/DRAM path with instruction buffering and a write-back data cache, and a fixed-latency on-chip memory path. The reported comparative evaluation uses the latter. Variable-latency memory remains a major limitation: the reported external-memory implementation reaches II=169.

The II=169 result is a property of where the variable latency was placed, not of the modeling method. In that implementation the AXI response is awaited inside the modeled iteration, so the memory round trip becomes part of the loop-carried recurrence and the scheduler must size the interval to the worst-case response rather than to the pipeline depth. The model already contains the mechanism needed to avoid this, because a miss is architecturally the same event as a load-use hazard: the consumer cannot proceed, the front end holds, a bubble enters ID/EX, and older work continues. Expressing the memory system as an additional synchronous process with its own persistent state, whose completion is one of the feedback conditions $H_t$ in Eq.~(1), keeps every modeled iteration a fixed-latency transition and turns a miss into a multi-iteration stall instead of a longer iteration. This is the same structure Comet uses to accommodate caches and multi-cycle operations. We have not built and measured that configuration, so the external-memory path remains the clearest limitation of the present evaluation.

\subsection{Consumption Timing and Memory Operations}
A value's consumption point determines which storage latency is legal. In the CPI-optimized pipeline, branches consume register operands during decode. A registered register-file read would make those operands arrive too late. The fully partitioned flip-flop implementation supports this timing, at substantial storage and multiplexing cost. Moving branch/jump resolution to execute lets the register-file read cross an explicit stage boundary; the same storage-binding request then becomes compatible with II=1. Two mirrored C++ arrays supply the distributed-RAM reads without hand-written RTL. The architectural price is a later redirect and increased branch penalty, rather than merely a change of storage directive.

Memory operations present a second example. Implementing a byte or halfword store as a word read, merge, and write adds a dependent access. Four independently addressed byte lanes eliminate that read-modify-write in the CPI-optimized design; the area-optimized design instead carries a byte-write strobe with a single-port write. Deferring host-interface acknowledgement also makes two otherwise competing stores mutually exclusive. These changes alter what must be scheduled, rather than asking the scheduler to ignore a real port conflict.

The designer therefore controls stage placement, validity, feedback selection, operand consumption, and memory access semantics. HLS controls the detailed schedule, internal registers, resource sharing, and legal bindings. Exploration alternates between structural changes and scheduling settings, with simulation and hardware measurement checking that the resulting architecture remains correct.

\begin{table*}[t]
    \caption{Measured throughput, all cores at 100~MHz}
    \begin{center}
    \footnotesize
    \begin{tabular}{|l|r|r|r|r|r|r|r|r|r|}
    \hline
    \textbf{Benchmark} & \textbf{Instr.} & \multicolumn{2}{c|}{\textbf{Pipelined, CPI-opt.}} & \multicolumn{2}{c|}{\textbf{Pipelined, area-opt.}} & \multicolumn{2}{c|}{\textbf{Ibex}} & \multicolumn{2}{c|}{\textbf{Algorithmic}} \\
    \cline{3-10}
     & & \textbf{cycles} & \textbf{cyc/instr} & \textbf{cycles} & \textbf{cyc/instr} & \textbf{cycles} & \textbf{cyc/instr} & \textbf{cycles} & \textbf{cyc/instr} \\
    \hline
    \texttt{rsort} & 190{,}632 & 204{,}506 & 1.073 & 210{,}127 & 1.102 & 272{,}603 & 1.430 & 762{,}563 & 4.000 \\
    \texttt{memcpy} & 29{,}614 & 32{,}154 & 1.086 & 34{,}675 & 1.171 & 50{,}676 & 1.711 & 118{,}491 & 4.001 \\
    \texttt{vvadd} & 3{,}963 & 4{,}442 & 1.121 & 4{,}905 & 1.237 & 6{,}409 & 1.616 & 15{,}887 & 4.008 \\
    \texttt{towers} & 4{,}118 & 4{,}620 & 1.122 & 4{,}853 & 1.178 & 7{,}223 & 1.753 & 16{,}507 & 4.008 \\
    \texttt{median} & 6{,}295 & 7{,}769 & 1.234 & 8{,}826 & 1.402 & 10{,}821 & 1.718 & 25{,}215 & 4.005 \\
    \texttt{dhrystone} & 298{,}743 & 380{,}571 & 1.274 & 427{,}878 & 1.432 & 483{,}350 & 1.618 & 1{,}195{,}015 & 4.000 \\
    \texttt{multiply} & 21{,}455 & 27{,}650 & 1.289 & 33{,}829 & 1.577 & 34{,}141 & 1.591 & 85{,}855 & 4.001 \\
    \texttt{qsort} & 134{,}812 & 180{,}627 & 1.340 & 199{,}335 & 1.479 & 221{,}242 & 1.641 & 539{,}283 & 4.000 \\
    \hline
    \textbf{Aggregate} & 689{,}632 & 842{,}339 & \textbf{1.221} & 924{,}428 & \textbf{1.340} & 1{,}086{,}465 & \textbf{1.575} & 2{,}758{,}816 & \textbf{4.000} \\
    \hline
    \end{tabular}
    \\[2pt]
    \begin{minipage}{\textwidth}\footnotesize
    Cycle counts are whole-program, measured by an external hardware counter, not the core's \texttt{mcycle}. \texttt{Instr.} is the CPI-optimized core's retired count; the four configurations agree to within 6 instructions on any benchmark and 12 in 689{,}632 overall, and each \texttt{cyc/instr} uses that configuration's own count. Rows are ordered by the CPI-optimized core's \texttt{cyc/instr}.
    \end{minipage}
    \label{tab:results}
    \end{center}
\end{table*}

\section{Experimental Evaluation}
\label{sec:experiments}
\subsection{Setup and Validation}
The algorithmic baseline, CPI-optimized model, area-optimized model, and Ibex reference execute identical binaries on a KCU116 board with a Kintex UltraScale+ \texttt{xcku5p-ffvb676-2-e}. Each sees 256 KB of single-cycle storage at the same access latency. HLS systems use separate instruction/data memories (130.5 BRAM tiles), while Ibex uses unified memory (66.5); these system-memory costs are excluded from the core-resource table. An external counter measures hardware cycles because the HLS \texttt{mcycle} advances per modeled iteration and would otherwise hide scheduling overhead.

Directed tests and eight kernels come from \texttt{riscv-tests}~\cite{riscv-tests}; unsupported instruction-cache-invalidation tests are excluded. Kernel sources are unchanged, but relinked with a minimal runtime replacing unsupported UART output with memory-based reporting and upstream verification checks. GCC 13.2.0 targets \texttt{rv32i\_zicsr}. CoreMark~\cite{coremark} runs 2500 iterations with reference CRCs checked. Reported validation includes RTL simulation and board execution; it is not an exhaustive proof of ISA or trap correctness.

Cycle counts are measured at 100 MHz. Execution time also uses each implementation's highest clock found with nonnegative post-route slack. Three candidate implementation flows are offered to every core, followed by frequency bisection; HLS project-mode strategies and Ibex non-project directives are comparable but not identical. The search resolution is 2 MHz for the baseline cores and 1 MHz for the area-optimized and final CPI-optimized cores. These are achieved clocks under the tested flows, not absolute physical limits. Tool release, exact source revision, and full scripts would be needed for complete reproduction.

A large language model (Claude Code) was used under the authors' direction to propose and implement candidate source transformations, including the resource-sharing, storage-binding and clock-target changes of Section~\ref{sec:area}, to generate separate directed tests and the build and measurement scripts, to run the synthesis and implementation sweeps, and to assist in analyzing the resulting data. The modeling methodology and the architectural decisions are the authors'; all reported data come from the tool flow and from board execution and were checked by the authors.

\begin{table*}[t]
    \caption{Measured execution time, each core at its own maximum clock}
    \begin{center}
    \footnotesize
    \begin{tabular}{|l|r|r|r|r|r|r|r|}
    \hline
    \textbf{Benchmark} & \textbf{Pipelined} & \textbf{Pipelined} & \textbf{Ibex} & \textbf{Algorithmic} & \textbf{CPI-opt.} & \textbf{Area-opt.} & \textbf{Algorithmic} \\
     & \textbf{CPI-opt.} & \textbf{area-opt.} & & & \textbf{/ Ibex} & \textbf{/ Ibex} & \textbf{/ Ibex} \\
     & \textbf{220~MHz} & \textbf{250~MHz} & \textbf{242~MHz} & \textbf{239~MHz} & & & \\
     & ($\mu$s) & ($\mu$s) & ($\mu$s) & ($\mu$s) & & & \\
    \hline
    \texttt{rsort} & 929.57 & 840.51 & 1126.46 & 3190.64 & $0.83\times$ & $0.75\times$ & $2.83\times$ \\
    \texttt{memcpy} & 146.15 & 138.70 & 209.40 & 495.78 & $0.70\times$ & $0.66\times$ & $2.37\times$ \\
    \texttt{vvadd} & 20.19 & 19.62 & 26.48 & 66.47 & $0.76\times$ & $0.74\times$ & $2.51\times$ \\
    \texttt{towers} & 21.00 & 19.41 & 29.85 & 69.07 & $0.70\times$ & $0.65\times$ & $2.31\times$ \\
    \texttt{median} & 35.31 & 35.30 & 44.71 & 105.50 & $0.79\times$ & $0.79\times$ & $2.36\times$ \\
    \texttt{dhrystone} & 1729.87 & 1711.51 & 1997.31 & 5000.06 & $0.87\times$ & $0.86\times$ & $2.50\times$ \\
    \texttt{multiply} & 125.68 & 135.32 & 141.08 & 359.23 & $0.89\times$ & $0.96\times$ & $2.55\times$ \\
    \texttt{qsort} & 821.03 & 797.34 & 914.22 & 2256.41 & $0.90\times$ & $0.87\times$ & $2.47\times$ \\
    \hline
    \textbf{Aggregate} & \textbf{3828.8} & \textbf{3697.7} & \textbf{4489.5} & \textbf{11543.2} & $\mathbf{0.85\times}$ & $\mathbf{0.82\times}$ & $\mathbf{2.57\times}$ \\
    \hline
    \end{tabular}
    \\[2pt]
    \begin{minipage}{\textwidth}\footnotesize
    Time is cycles divided by the achieved clock; cycle counts are those of Table~\ref{tab:results}. Each clock is the highest at which that design met timing post-route (WNS $\geq 0$) under the best of three implementation flows offered to every core. The area-optimized configuration is described in Section~\ref{sec:area}.
    \end{minipage}
    \label{tab:results_fmax}
    \end{center}
\end{table*}
\subsection{Does Explicit Structure Improve the Behavioral Baseline?}
\label{sec:modeling}
The algorithmic baseline implements one instruction per loop iteration without explicit pipeline registers, forwarding, or interlocks. Under a 4 ns target, register-file partitioning improves II from seven to four. The body occupies six scheduling states, so partial iteration overlap exists; this is not a strict absence of HLS pipelining. A forced aggressive schedule produces a 7.34 ns critical path instead of 2.85 ns and fails the target. The measured II=4 is consequently a property of this description and configuration, not a universal lower bound for behavioral HLS.

Table~\ref{tab:results} shows the CPI-optimized explicit-state design at 1.221 aggregate CPI versus 4.000 for the baseline and 1.575 for Ibex. Variation across kernels reflects hazards and redirects. The comparison supports the combined architectural reformulation, including hazard handling and memory changes; it does not isolate declaring registers alone. Ibex uses a different microarchitecture, so the 22.5\% cycle reduction is a case-study result rather than an intrinsic HLS advantage.

Table~\ref{tab:results_fmax} separates cycle efficiency from frequency. At 220 MHz, the CPI-optimized design takes $0.85\times$ Ibex's aggregate time despite Ibex reaching 242 MHz. The area-optimized configuration uses more cycles but reaches 250 MHz, yielding $0.82\times$ Ibex's time and the shorter absolute runtime of the two. Neither II nor clock frequency alone captures the tradeoff.

\subsection{What Remains Necessary to Reach II=1?}
\label{sec:opt}
Explicit state initially yields II=5, demonstrating that structural concurrency does not guarantee a one-cycle schedule. Deferring host-interface acknowledgement and removing registered boundaries around separately synthesized stage functions reduce II to two. Eliminating sub-word-store read-modify-write through byte lanes then reduces II to one. Five attempted changes---dual-port memory, removal of an already disabled atomic path, removal of \texttt{volatile}, writeback reordering, and counter-increment relocation---do not change II. These observations distinguish effective changes to scheduling structure from settings that leave its limiting depth intact.

Table~\ref{tab:ii} reports the memory transformation directly, measured on the core before the trap fix so that the memory organization is the only variable. Cycles approximately halve, but frequency falls from 300 to 226 MHz, giving a suite-time ratio of 0.664. Critical-path logic delay falls from 1.83 to 1.06 ns while routing delay rises from 1.34 to 3.02 ns. The later single-port strobe design also shows routing-dominated timing, so the measurements do not establish that byte-lane fanout alone causes the frequency loss. This is a measured physical tradeoff whose precise causal attribution remains open.

\begin{table}[t]
    \caption{Effect of the byte-lane data memory on the pipelined core, measured before the trap fix}
    \begin{center}
    \footnotesize
    \begin{tabular}{|l|r|r|r|}
    \hline
    \textbf{Benchmark} & \textbf{II${=}$2} & \textbf{II${=}$1} & \textbf{Ratio} \\
     & (cycles) & (cycles) & \\
    \hline
    \texttt{rsort} & 409{,}013 & 204{,}506 & 0.5000 \\
    \texttt{memcpy} & 64{,}309 & 32{,}154 & 0.5000 \\
    \texttt{vvadd} & 8{,}885 & 4{,}442 & 0.4999 \\
    \texttt{towers} & 9{,}241 & 4{,}620 & 0.4999 \\
    \texttt{median} & 15{,}539 & 7{,}769 & 0.5000 \\
    \texttt{dhrystone} & 761{,}143 & 380{,}571 & 0.5000 \\
    \texttt{multiply} & 55{,}301 & 27{,}650 & 0.5000 \\
    \texttt{qsort} & 361{,}255 & 180{,}627 & 0.5000 \\
    \hline
    \textbf{Aggregate} & \textbf{1{,}684{,}686} & \textbf{842{,}339} & \textbf{0.5000} \\
    \hline
    \textit{F}$_{\text{max}}$ (MHz) & 300 & 226 & 0.753 \\
    Suite time (ms) & 5.616 & 3.727 & 0.664 \\
    \hline
    \end{tabular}
    \\[2pt]
    \begin{minipage}{\columnwidth}\footnotesize
    Four byte lanes remove the read-modify-write behind every sub-word store, taking the loop from II=2 to II=1; the cycle ratio is $0.5$ on every benchmark. The restructuring costs clock speed, so the net gain in execution time is 34\%, not $2{\times}$. The two columns are successive versions of one core, both measured before the trap fix; the final CPI-optimized configuration in Tables~\ref{tab:results_fmax} and~\ref{tab:coremark} includes the fix and reaches 220~MHz.
    \end{minipage}
    \label{tab:ii}
    \end{center}
\end{table}
\subsection{How Do Structural Choices Change Area?}
\label{sec:area}
The CPI-optimized model uses 5,604 LUTs and 3,009 registers. Source consolidation replaces six shift expressions with one 33-bit arithmetic-right shifter and bit reversal for left shifts. Eleven adders and seventy-one comparators from separate expressions are consolidated into shared arithmetic/comparison paths; load/store address generation reuses the adder. Computing one CSR write value and enable replaces six write call sites. Together with unused decode removal, counter narrowing, a one-bit trap-target flag in place of a carried 32-bit address, removal of the register-file reset loop, and the single-port byte-strobe memory, these changes yield 2,936 LUTs with unchanged benchmark cycles. The reset loop is restored in the final configuration, where the register file is distributed RAM whose contents are undefined at power-up. Counter narrowing is a feature change and limits equivalence to the tested operating range; it should not be described as preserving all counter behavior.

Holding source fixed while relaxing an 8 ns scheduling target to 10 ns removes 452 registers without changing II or measured cycle behavior, but achieved frequency falls from 246 to 208 MHz. Targets of 6 ns or tighter lose II=1. The clock request therefore controls both scheduling storage and performance.

Moving branch/jump resolution to execute permits the distributed-RAM register file discussed above. In the isolated reported comparison it removes 601 LUTs and 746 registers, adds 40 distributed-RAM LUTs, and removes the fabric read multiplexers. Combined changes yield 1,980 LUTs and 1,595 registers at a 9 ns target; an 8 ns build uses 2,226 and 1,736 and reaches 256 MHz. The reported 65\% LUT and 47\% register reductions cost 9.7\% more kernel cycles and 6.6\% more CoreMark cycles. Hardware cycle counts match full-system RTL simulation on the measured kernels.

Table~\ref{tab:area} excludes memories and interconnect but includes the HLS host-control interface, absent from Ibex's scoped cell. Comet's published 2,032 LUTs and 1,503 registers use Catapult and Artix-7, with a different documented feature set. Its proximity to the area-optimized result is contextual; neither that row nor its Dhrystone CPI establishes a controlled superiority claim.

\begin{table}[t]
    \caption{Post-route resource usage, core logic only}
    \begin{center}
    \footnotesize
    \begin{tabular}{|l|r|r|r|}
    \hline
    \textbf{Core} & \textbf{LUT} & \textbf{FF} & \textbf{ MUX F7/F8} \\
    \hline
    \texttt{ibex} & 1{,}940 & 702 & 39 \\
    \texttt{algorithmic} & 6{,}327 & 3{,}140 & 95 \\
    \texttt{pipelined} & & & \\
    ~~\texttt{CPI-opt.} & 5{,}604 & 3{,}009 & 166 \\
    ~~\texttt{area-opt.} & 1{,}980 & 1{,}595 & 0 \\
    \texttt{comet}~\cite{comet_iccad}$^{\dagger\ddagger}$ & 2{,}032 & 1{,}503 & 260 \\
    \hline
    \end{tabular}
    \\[2pt]
    \begin{minipage}{\columnwidth}\footnotesize
    Scoped to the core instance alone: memories, BRAM controllers, the JTAG-to-AXI master, the interconnect, and the external cycle counter are excluded. No core uses block RAM or DSP; all memory is external. \texttt{F7/F8} counts the dedicated multiplexers used for selections too wide for a single LUT; narrower multiplexing is absorbed into LUT logic. Ibex and the area-optimized configuration hold their register files in distributed RAM; the other two use none. All four measured configurations implement \texttt{rv32i} with \texttt{Zicsr} and machine-mode traps, and the HLS cells additionally contain the AXI-lite interface through which the host starts the core, which Ibex's cell does not. $^{\dagger}$Comet is its published Artix-7 \texttt{xc7a12t} result~\cite{comet_iccad} from Catapult HLS and Vivado, on a different device generation; its \texttt{F7/F8} entry is that table's \texttt{Mux} column, which is not identified there as F7/F8 primitives, and its \texttt{rv32i} configuration is described as supporting ``only basic instructions'' with no mention of CSRs, privileged mode, or trap handling.
    \end{minipage}
    \label{tab:area}
    \end{center}
\end{table}
\begin{table}[t]
    \caption{CoreMark, ITERATIONS=2500}
    \begin{center}
    \footnotesize
    \begin{tabular}{|l|r|r|r|r|}
    \hline
    \textbf{Core} & \textbf{Cycles} & \textbf{CoreMark} & \textbf{Clock} & \textbf{CoreMark} \\
     & ($\times10^{6}$) & /MHz & (MHz) & (iter/s) \\
    \hline
    \texttt{pipelined} & & & & \\
    ~~\texttt{CPI-opt.} & 2{,}573 & 0.971 & 220 & 213.7 \\
    ~~\texttt{area-opt.} & 2{,}744 & 0.911 & 250 & 227.8 \\
    \texttt{ibex} & 2{,}734 & 0.915 & 242 & 221.3 \\
    \texttt{algorithmic} & 7{,}466 & 0.335 & 239 & 80.0 \\
    \hline
    \end{tabular}
    \\[2pt]
    \begin{minipage}{\columnwidth}\footnotesize
    All runs exceed CoreMark's 10~s minimum and report the four reference CRCs. \texttt{CoreMark/MHz} is clock-independent: identical at 100~MHz and at \textit{F}$_{\text{max}}$.
    \end{minipage}
    \label{tab:coremark}
    \end{center}
\end{table}
\subsection{Workload Dependence and Limits}
Table~\ref{tab:coremark} shows why the conclusions must remain workload-specific. The CPI-optimized design leads Ibex in CoreMark/MHz (0.971 versus 0.915), but trails it in iterations/s (213.7 versus 221.3), while the area-optimized design reaches 227.8 iterations/s despite a lower normalized score of 0.911. The eight-kernel results and CoreMark consistently motivate reporting area, cycles, and achieved frequency together, but do not cover caches, external-memory stalls, or broader application mixes. Generality is demonstrated conceptually through the state-transition formulation and experimentally on one processor family; non-processor evaluation and a matched RTL architecture remain future work.

\section{Conclusion}
Explicit architectural modeling separates the state, communication, and hazard decisions that make concurrency legal from the scheduling and binding decisions HLS can optimize. The RV32I example shows how communication paths, consumption timing, and memory-operation structure determine II and performance--area tradeoffs. The measured configurations demonstrate competitive case-study performance and substantial source-driven area reductions, while differing architectures, memory systems, and tool flows limit broader comparisons. The same steps apply elsewhere: name the persistent state, place the process boundaries, and specify which version of each shared value a consumer reads, when it is consumed, and which ports it requires. The method suits designs whose next state depends on an arbitration or allocation decision: a switch arbiter, a cryptographic round and key schedule, or a cache controller where a miss is a hold condition rather than a longer iteration, none of which we have evaluated here.
\bibliographystyle{ACM-Reference-Format}
\bibliography{HLS_references}

@inproceedings{comet_iccad,
  author    = {Rokicki, Simon and Paturel, Joseph and Sentieys, Olivier},
  title     = {What You Simulate Is What You Synthesize: Designing a Processor Core from {C++} Specifications},
  booktitle = {2019 IEEE/ACM International Conference on Computer-Aided Design (ICCAD)},
  year      = {2019},
  doi = {10.1109/ICCAD45719.2019.8942177}
}

@misc{comet,
  author       = {Rokicki, Simon and Paturel, Joseph and Sentieys, Olivier},
  title        = {Comet: a {RISC-V} Core Synthesized from {C++} Specifications},
  year         = {2022},
  note         = {Poster, RISC-V Week, Paris},
  howpublished = {\url{https://inria.hal.science/hal-03885663}}
}

@inproceedings{hl5,
  author    = {Mantovani, Paolo and Guglielmo, Giuseppe Di and Carloni, Luca P.},
  title     = {{HL5}: A 32-bit {RISC-V} Processor Designed with High-Level Synthesis},
  booktitle = {2020 IEEE Custom Integrated Circuits Conference (CICC)},
  year      = {2020},
  doi = {10.1109/CICC48029.2020.9075913}

}

@incollection{hl5_scaling,
  author    = {Schoder, Johannes and B{\"u}cker, H. Martin},
  title     = {Scaling an Augmented {RISC-V} Processor Design with High-Level Synthesis},
  booktitle = {Embedded Computer Systems: Architectures, Modeling, and Simulation},
  publisher = {Springer},
  year      = {2024},
  doi = {10.1007/978-3-031-73716-9_22}
}

@inproceedings{toker_rv32i,
  author    = {Toker, Onur},
  title     = {A High-Level Synthesis Approach for a {RISC-V} {RV32I}-Based System on Chip and Its {FPGA} Implementation Using Open Source Tools},
  booktitle = {Proceedings of the 10th International Electronic Conference on Sensors and Applications},
  publisher = {MDPI},
  year      = {2023},
  doi   = {10.3390/ecsa-10-16212}
}

@techreport{rocketchip,
  author      = {Asanovi\'{c}, Krste and Avizienis, Rimas and Bachrach, Jonathan and others},
  title       = {The Rocket Chip Generator},
  institution = {EECS Department, University of California, Berkeley},
  year        = {2016},
  number      = {UCB/EECS-2016-17}
}

@inproceedings{riscv_survey,
  author    = {D{\"o}rflinger, Alexander and Albers, Mark and Kleinbeck, Benedikt and Guan, Yejun and Michalik, Harald and Klink, Raphael and Blochwitz, Christopher and Nechi, Anouar and Berekovic, Mladen},
  title     = {A Comparative Survey of Open-Source Application-Class {RISC-V} Processor Implementations},
  booktitle = {Proceedings of the 18th ACM International Conference on Computing Frontiers (CF '21)},
  publisher = {ACM},
  address   = {New York, NY, USA},
  year      = {2021},
  pages     = {12--20},
  doi       = {10.1145/3457388.3458657}
}

@article{pragma_nlp,
  author  = {Pouget, St{\'e}phane and Pouchet, Louis-No{\"e}l and Cong, Jason},
  title   = {Automatic Hardware Pragma Insertion in High-Level Synthesis: A Non-Linear Programming Approach},
  journal = {ACM Transactions on Design Automation of Electronic Systems},
  number = {2},
  volume  = {30},
  articleno = {26}, 
  numpages = {44},
  year    = {2025},
  doi     = {10.1145/3711847}
}

@article{autoscaledse,
  author  = {Jun, HyeGang and Ye, Hanchen and Jeong, Hyunmin and Chen, Deming},
  title   = {{AutoScaleDSE}: A Scalable Design Space Exploration Engine for High-Level Synthesis},
  journal = {ACM Transactions on Reconfigurable Technology and Systems},
  number = {3},
  volume  = {16},
  articleno = {46}, 
  numpages = {30},
  year    = {2023},
  doi     = {10.1145/3572959}
}

@article{allo,
  author  = {Chen, Hongzheng and Zhang, Niansong and Xiang, Shaojie and Zeng, Zhichen and Dai, Mengjia and Zhang, Zhiru},
  title   = {Allo: A Programming Model for Composable Accelerator Design},
  journal = {Proceedings of the ACM on Programming Languages},
  number = {PLDI},
  volume  = {8},
  articleno = {171},
  pages = {593--620},
  year    = {2024},
  doi     = {10.1145/3656401}
}

@inproceedings{hida,
  author    = {Ye, Hanchen and Jun, Hyegang and Chen, Deming},
  title     = {{HIDA}: A Hierarchical Dataflow Compiler for High-Level Synthesis},
  booktitle = {Proceedings of the 29th ACM International Conference on Architectural Support for Programming Languages and Operating Systems (ASPLOS)},
  year      = {2024},
  doi       = {10.1145/3617232.3624850}
}

@inproceedings{streamhls,
  author    = {Basalama, Suhail and Cong, Jason},
  title     = {Stream-{HLS}: Towards Automatic Dataflow Acceleration},
  booktitle = {Proceedings of the 2025 ACM/SIGDA International Symposium on Field Programmable Gate Arrays (FPGA)},
  year      = {2025},
  doi       = {10.1145/3706628.3708878}
}

@inproceedings{sisyphus,
  author    = {Pouget, St{\'e}phane and Pouchet, Louis-No{\"e}l and Cong, Jason},
  title     = {A Unified Framework for Automated Code Transformation and Pragma Insertion},
  booktitle = {Proceedings of the 2025 ACM/SIGDA International Symposium on Field Programmable Gate Arrays (FPGA)},
  year      = {2025},
  doi       = {10.1145/3706628.3708873}
}

@article{prometheus,
  author  = {Pouget, St{\'e}phane and Lo, Michael and Pouchet, Louis-No{\"e}l and Cong, Jason},
  title   = {Holistic Optimization Framework for {FPGA} Accelerators},
  journal = {ACM Transactions on Design Automation of Electronic Systems},
  number = {1},
  volume  = {31},
  articleno = {7},
  numpages = {37},
  year    = {2025},
  doi     = {10.1145/3769307}
}

@manual{sifive-e31,
  author = {{SiFive, Inc.}},
  title  = {{SiFive E31 Core Complex Manual v2p0}},
  year   = {2019},
  url    = {https://static.dev.sifive.com/SiFive-E31-Manual-v2p0.pdf}
}

@misc{ibex,
  author       = {{lowRISC}},
  title        = {{Ibex RISC-V Core}},
  year         = {2024},
  howpublished = {\url{https://github.com/lowRISC/ibex}}
}

@misc{opentitan,
  author       = {{lowRISC}},
  title        = {{OpenTitan: Open Source Silicon Root of Trust}},
  year         = {2024},
  howpublished = {\url{https://opentitan.org}}
}

@misc{picorv32,
  author       = {Wolf, Claire},
  title        = {{PicoRV32: A Size-Optimized RISC-V CPU}},
  year         = {2024},
  howpublished = {\url{https://github.com/YosysHQ/picorv32}}
}

@misc{vexriscv,
  author       = {{SpinalHDL}},
  title        = {{VexRiscv}},
  year         = {2024},
  howpublished = {\url{https://github.com/SpinalHDL/VexRiscv}}
}

@misc{riscv-tests,
  author       = {{RISC-V International}},
  title        = {{riscv-tests}},
  year         = {2024},
  howpublished = {\url{https://github.com/riscv-software-src/riscv-tests}} 
}

@misc{coremark,
  author       = {{EEMBC}},
  title        = {{CoreMark: An EEMBC Benchmark}},
  year         = {2024},
  howpublished = {\url{https://www.eembc.org/coremark/}}
}
\end{document}